\documentclass{amsart}
\input{diagxy}

\usepackage{amssymb}
\newtheorem{theorem}{Theorem}[section]
\theoremstyle{definition}
\newtheorem{definition}[theorem]{Definition}
\def\i{{\rm i}}
\newcommand{\cCas}{\mathbf{Cal^0_{Sym}}}
\newcommand{\cC}{\mathbf{C}}
\newcommand{\cc}{\mathbf{c}}
\newcommand{\op}{{op}}
\renewcommand{\l}{\mathcal{L}}
\newcommand{\h}{\mathcal{H}}
\newcommand{\Sep}{\mathcal{S}}
\newcommand{\sep}{\mathsf{S}}
\newcommand{\unit}{\mathsf{U}}
\newcommand{\pro}{\mathsf{P}}
\newcommand{\aut}{\mathsf{Aut}}
\def\prs{\s_{_\Downarrow}}
\def\s{\Sigma}
\def\t{\times}
\def\p{\perp}
\def\ot{\otimes}
\def\aerts{\,{\land\kern-.85em\bigcirc}}
\def\otf{\,{\vee\kern-.85em\bigcirc}}
\def\ots{\,{\Downarrow\kern-1.08em\bigcirc}}
\begin{document}
\title{Orthocomplementation and compound systems}
\author{Boris Ischi}
\address{Boris Ischi, Laboratoire de Physique des Solides,
Universit\'e Paris-Sud, B\^atiment 510, 91405 Orsay, France}
\email{ischi@kalymnos.unige.ch}
\thanks{Supported by the Swiss National Science Foundation.}
\begin{abstract}
In their 1936 founding paper on quantum logic, Birkhoff and von
Neumann postulated that the lattice describing the experimental
propositions concerning a quantum system is orthocomplemented. We
prove that this postulate fails for the lattice $\l_{sep}$
describing a compound system consisting of so called {\it
separated} quantum systems. By separated we mean two systems
prepared in different ``rooms'' of the lab, and before any
interaction takes place. In that case the state of the compound
system is necessarily a product state. As a consequence, Dirac's
superposition principle fails, and therefore $\l_{sep}$ cannot
satisfy all Piron's axioms. In previous works, assuming that
$\l_{sep}$ is orthocomplemented, it was argued that $\l_{sep}$ is
not orthomodular and fails to have the covering property. Here we
prove that $\l_{sep}$ cannot admit and orthocomplementation.
Moreover, we propose a natural model for $\l_{sep}$ which has the
covering property.

\end{abstract}
\maketitle
\section{Introduction}

A cornerstone in physics is the concept of a mathematical
phase-space $\s_S$ associated with a physical system $S$,
representing all possible states of $S$. For instance, a classical
particle is at each instant $t$ associated with a point
$(\vec{x}_t,\vec{p}_t)\in\mathbb{R}^6$ where $\vec{x}_t$ and
$\vec{p}_t$ are the position and the momentum of the particle at
time $t$ respectively. On the other hand, in quantum theory, it is
assumed that there is a complex Hilbert space $\h_S$ associated
with $S$, such that $\s_S=(\h_S-0)/\mathbb{C}$, the set of
one-dimensional subspaces of $\h_S$ \cite{vonNeumann:handbook}.

In \cite{Birkhoff/Neumann:1936}, \S 2, Birkhoff and von Neumann
call a measurement $\mathcal{M}$ on a physical system $S$,
together with a given subset $\sigma\subseteq \mathcal{O_M}$ of
possible outcomes, an {\it experimental proposition} concerning
the system $S$. Experimental propositions can be correlated with
subsets of $\s_S$ by assigning to each proposition $P$, the set
$\mu(P)$ of states in which the measurement yields with certainty
an outcome in $\sigma$. In the sequel, we shall denote the image
of the map $\mu$, ordered by set-inclusion, by $\l_S$, and call it
the {\it property poset} of $S$. Note that $\l_S\subseteq
2^{\s_S}$, and obviously $\emptyset$ and $\s_S$ are in $\l_S$.

In classical mechanics $\mu(\neg P)=\s\backslash\mu(P)$, where
$\neg P$ denotes the proposition defined by the same measurement,
but with the complementary subset of outcomes
$\mathcal{O_M}\backslash\sigma$. This means that for any state,
the probability that the outcome lies in $\sigma$ (respectively in
$\mathcal{O_M}\backslash\sigma$) is either $1$ or $0$. Hence, in
classical mechanics, $\l_S$ is a suborthoposet of $2^{\s_S}$.

For quantum theory the situation is totally different: The
measurement $\mathcal{M}$ is associated with a self-adjoint
operator on $\h_S$, and $\sigma$ with a projector $\mathsf{P}_V$
on a closed subspace $V$ of $\h_S$. For a given state $p$, the
probability that the outcome of $\mathcal{M}$ lies in $\sigma$
(respectively in $\mathcal{O_M}\backslash\sigma$) is given by
$\Vert \mathsf{P}_V(\phi)\Vert^2$ (respectively by $\Vert
\mathsf{P}_{V^\p}(\phi)\Vert^2$) where $\phi\in p$ with
$\Vert\phi\Vert=1$. Whence, $\mu(P)=(V-0)/\mathbb{C}$ and
$\mu(\neg P)=(V^\p-0)/\mathbb{C}$. Therefore, in quantum theory,
$\l_S$ is a suborthoposet of
$\pro(\h_S)=\{(V-0)/\mathbb{C}\,;\,V\subseteq\h_S,\,V^{\p\p}=V\}$,
the lattice of closed subspaces of $\h_S$.

For both classical mechanics and quantum theory, Birkhoff and von
Neumann postulated that $\l_S$ is an orthocomplemented lattice
(see \cite{Birkhoff/Neumann:1936}, \S 5-6). More precisely, $\l_S$
is assumed to be a subortholattice, of $2^{\s_S}$ in the classical
case, and of $\pro(\h_S)$ in the quantum case.

In this paper, we want to study the mathematical structure of the
phase-space $\s_S$ and the property poset $\l_S$ of a compound
system $S$ consisting of two {\it separated} quantum systems $S_1$
and $S_2$. By separated, we mean two systems (electrons, atoms or
whatever) prepared in two different ``rooms'' of the lab, and
before any interaction takes place. In that case, we denote $\s_S$
by $\s_{sep}$, and $\l_S$ by $\l_{sep}$. As a main result, we show
that $\l_{sep}$ cannot admit an orthocomplementation.

What do we know about $\s_{sep}$ and $\l_{sep}$? In quantum
theory, the phase-space of a two-body system is given by
$(\h_1\ot\h_2-0)/\mathbb{C}$, hence the state of $S$ can be either
entangled or a product state \cite{vonNeumann:handbook}. Entangled
states have been observed in many experiments, involving pairs of
photons (see \cite{AspectNature:1999} and references herein) or
massive particles \cite{RoweNature:2001}. Gisin proved that any
entangled state violates a Bell inequality \cite{Gisin:1991}.
Therefore, for separated systems as defined here, the state is
necessarily a product $p_1\ot p_2$ with
$p_i\in(\h_i-0)/\mathbb{C}$. Whether the two systems are fermions
or bosons does not matter. Since they are prepared independently
and do not interact, they are distinguishable and not correlated.
As a consequence, we can put $\s_{sep}=\s_{S_1}\t\s_{S_2}$.

Further, let $P_1$ and $P_2$ be experimental propositions
concerning $S_1$ and $S_2$ respectively. Then, obviously, both
$P_1$ and $P_2$ are also experimental propositions concerning the
compound system $S$. Moreover,
\[\mu(P_1)=\mu_1(P_1)\t\s_2\ \mbox{and}\ \mu(P_2)=\s_1\t\mu_2(P_2)\, .\]
Now, since $S_1$ and $S_2$ are totally independent from each
other, we can perform $P_1$ and $P_2$ simultaneously (or one after
the other) and define the experimental propositions concerning the
compound system $P_1\,\mbox{AND}\,P_2$ and $P_1\,\mbox{OR}\,P_2$.
Then, obviously
\[\begin{split}
\mu(P_1\, \mbox{AND}\,P_2)&=\mu_1(P_1)\t\mu_2(P_2)\\
\mu(P_1\,\mbox{OR}\,P_2)&=\mu_1(P_1)\t\s_2\cup \s_1\t\mu_2(P_2)\,
.\end{split}\]
Note that if we only consider those kind of experimental
propositions on the compound system $S$, then $\l_{sep}$ is given
by the {\it separated product} of Aerts $\l_{S_1}\aerts\l_{S_2}$
defined in \cite{Aerts:1982} (see Section \ref{SectionResults}).

In Section \ref{SectionArgumentsAgainstSeparatedProduct}, we will
see that some important experimental propositions are not
described by the separated product of Aerts. This means that
$\l_{sep}$ cannot be constructed by simply considering the
conjunctions and disjunctions of propositions concerning $S_1$ and
$S_2$. As a consequence, in order to investigate the mathematical
structure of $\l_{sep}$, we proceed as follows. First, we show
that $\l_{sep}$ is a {\it weak tensor product} of $\l_{S_1}$ and
$\l_{S_2}$ (Section \ref{SectionPhysicalHypotheses}), and then, we
prove that if a weak tensor product admits an
orthocomplementation, then it is isomorphic to the separated
product of Aerts; whence follows our main claim, namely that
$\l_{sep}$ cannot admit an orthocomplementation. We end the paper
with some open questions concerning weak tensor products in the
last Section \ref{SectionOpenQuestion}.
\section{Two arguments against the separated
product}\label{SectionArgumentsAgainstSeparatedProduct}
\subsection{Missing properties}\label{SubSectionMissingProperties}

Let $S$ be any physical system undergoing some time evolution from
a time $t_0$ to a time $t_1$. Let $U:{\s_S}_{t_0}\rightarrow
{\s_S}_{t_1}$ be a map describing this time evolution. Let
$\mathcal{M}_{t_1}$ be a measurement which can be performed on $S$
at time $t_1$, and let $P_{t_1}$ be an experimental proposition
associated with $\mathcal{M}_{t_1}$. Then, Daniel pointed out that
we can define an experimental proposition $\Phi(P_{t_1})$
concerning $S$ at time $t_0$, by the prescription: ``Let $S$
evolve from time $t_0$ to time $t_1$ and perform
$\mathcal{M}_{t_1}$''; obviously,
$U^{-1}(\mu(P_{t_1}))=\mu(\Phi(P_{t_1}))$ \cite{Daniel:1989}. As a
consequence, if we ask ${\l_S}_{t_0}$ to describe also those kind
of experimental propositions, then for any $b\in{\l_S}_{t_1}$,
$U^{-1}(b)\in{\l_S}_{t_0}$.

Note that in quantum theory, the time evolution of an isolated
system $S$ is described by a unitary operator on the Hilbert space
$\h_S$; moreover unitary operators preserve closed subspaces. In
general, it seems natural to require that the property poset of a
physical system, describes all experimental propositions defined
by Daniel's prescription, applied to any possible time evolution.
Is it true for the separated product? To answer this question, we
must first know what kind of time evolutions can undergo two
initially separated quantum systems $S_1$ and $S_2$.

\begin{figure}[t]
\setlength{\unitlength}{1cm} \frame{
  \begin{picture}(12,5.7)(0,0)
    \put(-1,0){\includegraphics{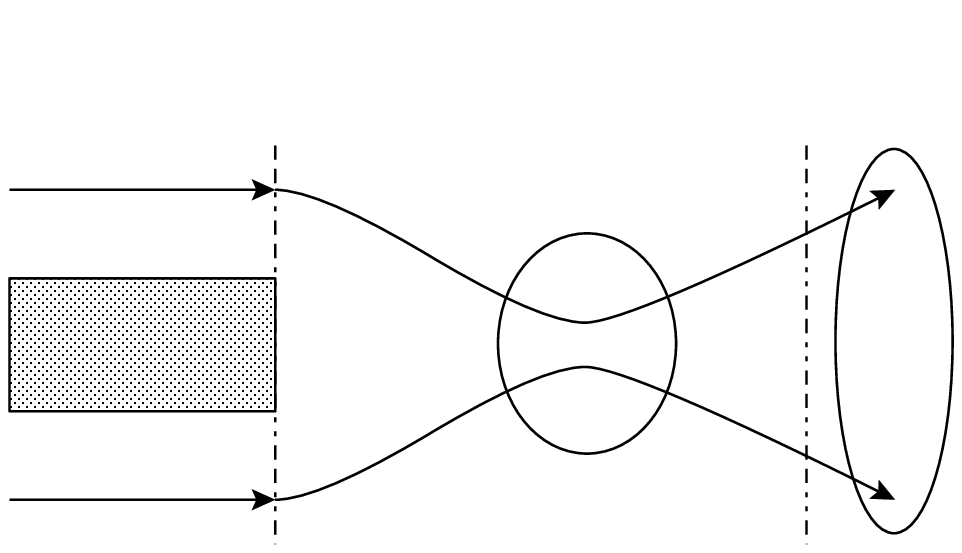}}
    \put(1,5){\makebox(0,0){$S_1$}}
    \put(1,1.7){\makebox(0,0){$S_2$}}
    \put(2.5,1){\makebox(0,0){Separated}}
    \put(4,1){\makebox(0,0){$t_0$}}
    \put(9.5,1){\makebox(0,0){$t_1$}}
    \put(7.2,1){\makebox(0,0){Interaction}}
    \put(10.8,1){\makebox(0,0){Measurement}}
  \end{picture}
}
\caption{} \label{FigureExperiment}
\end{figure}

First, on can simply keep each system in its own ``room'', and let
the systems evolve. Consider now the experimental situation
represented schematically in Figure \ref{FigureExperiment}. Two
quantum systems $S_1$ and $S_2$ are prepared in two different
``rooms'' of the lab and stay in their own ``room'' until a time
$t_0$. Then some interaction is ``switched on'', and finally a
measurement is performed at some later time $t_1$, after the
interaction has taken place. This is typically a situation
encountered in scattering experiments.

According to quantum theory, the time evolution from $t_0$ to
$t_1$ is given by a unitary operator $U=\exp[-\i (t_1-t_0)(H_1\ot
1+1\ot H_2 +V)]$, where $H_i$ is the free Hamiltonian acting on
$\h_i$, and $V$ the interaction. As discussed in the introduction,
at time $t_0$ the sate of $S$ is a product state. Now, because of
the interaction $V$, $U$ transforms instantaneously initial
product states into entangled states. This means that the
mathematical structure of the property poset $\l_S$ of the
compound system after time $t_0$ is very different from that of
$\l_{sep}$. How this change occurs is far behind the scope of this
article. However, we can expect that at time $t_1$, experimental
propositions concerning the compound system correspond to closed
subspaces of $\h_1\ot\h_2$. Hence, according to Daniel's
principle, it is natural to ask that for all
$b\in\pro(\h_1\ot\h_2)$, the set of product states contained in
$U^{-1}(b)$ is an element of $\l_{sep}$. This is not true for the
separated product (see \cite{Ischi:2004}, Theorem 10.4).

To make the discussion more definite, let us consider the set-up
proposed by Tanamoto in \cite{Tanamoto:2000} to realize a
solid-state quantum C-NOT gate. In this example, the systems $S_1$
and $S_2$ are electrons. Each electron is injected at time $t_0$
in a double quantum dot. The complete localization of the electron
in the first or the second dot is denoted by $0=(1,0)$ and
$1=(0,1)$ respectively. At low enough temperatures, only the
combinations of these two states need to be taken into account,
hence we can assume that $\h_{S_i}=\mathbb{C}^2$ (qubits). The two
electrons are injected from two different gates, each of them in
its own double quantum dot. Therefore, at time $t_0$ the two
electrons are separated. The evolution operator of the quantum
C-NOT gate (say from time $t_0$ to time $t_1$) is given by
$\mathcal{U}=\frac{1}{2}(1-\sigma_z)\ot\sigma_x+
\frac{1}{2}(1+\sigma_z)\ot\mathsf{id}$, where $\sigma_z$ and
$\sigma_x$ denote the Pauli's matrices. Note that
$\mathcal{U}^2=\mathcal{U}$ and that $\mathcal{U}$ sends the
(non-normalized) state $(0+1)\ot 0$ to the entangled state $0\ot
0+1\ot 1$ \cite{BarenkoPRL:1995}. As a consequence, if we assume
for instance that the subspace of $\mathbb{C}^2\ot\mathbb{C}^2$
orthogonal to $(0+1)\ot 0$ corresponds to an experimental
proposition concerning the compound system at time $t_1$, then
according to Daniel's principle, the set of product states
contained in $(0\ot 0+1\ot 1)^\p$ must be an element of
${\l_S}_{t_0}=\l_{sep}$. Now, it is an easy exercise to check that
$\pro(\mathbb{C}^2)\aerts\pro(\mathbb{C}^2)$ does not contain this
particular subset of product states.

In sum, the heuristic arguments given above, seem to indicate that
some important experimental propositions are not described by the
separated product of Aerts. Let us give a second argument against
the separated product.
\subsection{Propensities}

It is natural to assume the existence of a {\it propensity} map
$\omega:\s_{sep}\t\l_{sep}\rightarrow[0,1]$ as defined for
instance in \cite{Gisin:1984}. By a result of Pool, every
orthocomplemented lattice which admits a propensity map is
orthomodular \cite{Pool:1968}. Suppose that $\l_{S_1}=\pro(\h_1)$,
$\l_{S_2}=\pro(\h_2)$, and that
$\l_{sep}=\pro(\h_1)\aerts\pro(\h_2)$. Note that
$\pro(\h_1)\aerts\pro(\h_2)$ is a complete atomistic
orthocomplemented lattice. Then, $\l_{sep}$ cannot admit a
propensity map, since by Aerts's theorem, if $\l_1$ and $\l_2$ are
complete atomistic orthocomplemented lattices and $\l_1\aerts\l_2$
is orthomodular, then $\l_1$ or $\l_2$ is distributive
\cite{Aerts:1982}.
\section{Physical hypotheses}\label{SectionPhysicalHypotheses}
\subsection{General assumptions on the property posets}

In the sequel, for any physical system $S$, we shall consider not
only experimental propositions, but more generally all
$\{0,1\}$-valued experiments on $S$. Hence, following Piron
\cite{Piron:handbook} and Aerts \cite{Aerts:1982}, we shall assume
that $\l_S$ is closed under arbitrary set-intersections ({\it
i.e.} $\cap\omega\in\l_S$, for all $\omega\subseteq\l_S$). Let us
repeat the physical argument. Let $\{\mu(\alpha_i)\in\l_S\}_{i\in
I}$ with $\alpha_i$ $\{0,1\}$-valued experiments on $S$. Define
$\pi_i\alpha_i$ by the prescription: ``perform any $\alpha_i$''.
Then obviously, $\mu(\pi_i\alpha_i)=\cap_i\mu(\alpha_i)$.

We make a second general hypothesis on property posets. For
$p\in\s_S$, let $\varepsilon_p$ denote all $\{0,1\}$-valued
experiments $\alpha$ on $S$, such that $p\in\mu(\alpha)$. Then,
following Aerts \cite{Aerts:1982}, we assume that for any two
states $p$ and $q$ in $\s_S$,
$\varepsilon_p\nsubseteq\varepsilon_q$. Whence follows that
$\{p\}=\cap\{\mu(\alpha)\,;\,\alpha\in\varepsilon_p\}$, for all
$p\in\s_S$.

As a consequence, $\l_S$ contains $\emptyset$, $\s_S$, and all
singletons of $\s_S$, and $\l_S$ is closed under arbitrary
set-intersections. Hence, $\l_S$ is the set of closed subspaces of
a simple closure space.

To be short, we call a set $\l$ of subsets of a nonempty set $\s$,
closed under arbitrary set-intersections, and containing
$\emptyset$, $\s$, and all singletons of $\s$, a {\it simple
closure space} on $\s$. Note that a simple closure space is a
complete atomistic lattice. Moreover, if $\l$ is a complete
atomistic lattice, then $\{\s[a]\,;\,a\in\l\}$, where $\s[a]$
denotes the set of atoms under $a$, is a simple closure space on
the set of atoms of $\l$.

Four our main result, we need to assume moreover that $\l_{S_1}$
and $\l_{S_2}$ are orthocomplemented with the covering property,
which is of course true if $\l_{S_i}=\pro(\h_i)$ with $\h_i$ a
complex Hilbert space.
\subsection{Assumptions relating $\l_{S_i}$ and $\l_{S_{sep}}$}

We assume that $\l_{sep}$ is a {\it weak tensor product} of
$\l_{S_1}$ and $\l_{S_2}$:

\begin{definition} Let $\l_1\subseteq 2^{\s_1}$ and
$\l_2\subseteq 2^{\s_2}$ be simple closure spaces on $\s_1$ and
$\s_2$ respectively. Then $\sep(\l_1,\l_2)$ is defined as the set
of all simple closure spaces $\l\subseteq 2^\s$ on $\s$ such that
\begin{enumerate}
\item[P1] $\s=\s_1\t\s_2$,
\item[P2] $a_1\t\s_2\cup\s_1\t a_2\in\l$, $\forall a_1\in\l_1$,
$a_2\in\l_2$,
\item[P3] $\forall p_i\in\s_i,\, A_i\subseteq\s_i$, $[p_1\t
A_2\in\l\Rightarrow A_2\in\l_2]$ and $[A_1\t p_2\in\l\Rightarrow
A_1\in\l_1]$.
\end{enumerate}
We call elements of $\sep(\l_1,\l_2)$ weak tensor products of
$\l_1$ and $\l_2$. Let $T_i\subseteq\aut(\l_i)$, the group of
automorphisms of $\l_i$ ({\it i.e.} bijective maps preserving all
meets and joins). Then we define $\Sep_{_{T_1T_2}}(\l_1,\l_2)$ as
the subset of all $\l\in\sep(\l_1,\l_2)$ such that
\begin{enumerate}
\item[P4] $\forall v_i\in T_i$, $\exists u\in\aut(\l)$ $\vert$
$u(p_1,p_2)=(v_1(p_1),v_2(p_2))$, $\forall (p_1,p_2)\in\s$.
\end{enumerate}
\end{definition}

Note that for a simple closure space $\l\subseteq 2^\s$ on $\s$,
we omit the brackets when writing singletons and call elements of
$\s$ atoms. Moreover, for $u\in\aut(\l)$, we also write $u$ for
the bijective map on $\s$ induced by $u$.

For our main result, we only need Axioms P1-P3. Axioms P1 and P2
have already been discussed in the introduction.

\noindent{\bf Axiom P4} If $S_1$ and $S_2$ are quantum systems
described by two complex Hilbert spaces $\h_1$ and $\h_2$, then it
is indeed natural to assume that Axiom P4 holds for
$(T_1,T_2)=(\unit(\h_1),\unit(\h_i))$, where $\unit(\h_i)$ denotes
the group of automorphisms of $\pro(\h_i)$ induced by unitary
maps. In words, it is natural to assume that products of unitary
maps represent physical symmetries of the compound system. Of
course, we can expect that Axiom P4 also holds for pairs of
antiunitary maps. Suppose now that the automorphisms in $T_1$ and
$T_2$ describe possible time evolutions of each system. Then,
according to the discussion in Section
\ref{SubSectionMissingProperties}, Axiom P4 must hold for $T_1$
and $T_2$.

\noindent{\bf Axiom P3} From the experimental standpoint, the
system $S_1$ can certainly not be prepared in any given state.
However, we can reasonably assume that there is at least one state
(say $p_0$) in which $S_1$ can be prepared, whatever the system
$S_1$ might be. Now, suppose that $p_0\t B\in\l_{sep}$ and let
$P\in\mathcal{P}_{sep}$ such that $\mu(P)=p_0\t B$. Define a
$\{0,1\}-$valued experiment $P_2$ as ``prepare system $S_1$ in
room 1 in the state $p_0$ and perform P''. Then obviously, $P_2$
is a $\{0,1\}-$valued experiment on $S_2$, and $\mu_2(P_2)=B$,
hence $B\in\l_2$. Therefore, Axiom P3 follows from Axiom P4, if
$T_1$ and $T_2$ act transitively on $\s_{S_1}$ and $\s_{S_2}$
respectively, and contain the identity, which is of course true if
$T_1=\unit(\h_1)$ and $T_2=\unit(\h_2)$. It is important to note,
that to justify Axiom P3, we need to assume both the existence for
each system of a particular state in which each it can be
prepared, and enough physical symmetries. Indeed, if for instance
$T_1$ corresponds to automorphisms describing possible time
evolutions of $S_1$, then $T_1$ certainly does not act
transitively on $\s_{S_1}$. Finally, not that if the system $S_1$
can be prepared in a given state $p_0$, and if $U_1$ is a time
evolution sending the initial state $p_0$ to a final state $p_1$,
this does not mean that $S_1$ can be prepared in the state $p_1$.
\section{Mathematical results}\label{SectionResults}

Before we present some mathematical results concerning weak tensor
products, we want to emphasize on the fact that obviously, a weak
tensor product (hence $\l_{sep}$) cannot be isomorphic to the
lattice of closed subspaces of a Hilbert space. Therefore some of
Piron's axioms must be failing in $\l_{sep}$ \cite{Piron:1964}. In
the light of Theorems 34.5 and 33.7 in
\cite{Maeda/Maeda:handbook}, one can expect that $\l_{sep}$ is
neither orthocomplemented with the covering property, nor a
DAC-lattice ($\l$ is a DAC-lattice if both $\l$ and the dual of
$\l$ are atomistic with the covering property).
\subsection{Generalities}

\begin{definition} Let $\l_1\subseteq 2^{\s_1}$ and
$\l_2\subseteq2^{\s_2}$ be simple closure spaces on $\s_1$ and
$\s_2$ respectively. Then
\[
\begin{split}
\l_1\aerts\l_2&:=\{\cap\omega\,;\,\omega\subseteq\{a_1\t\s_2\cup\s_1\t
a_2\,;\,a_1\in\l_1,\,a_2\in\l_2\}\}\, ,\\
\l_1\otf\l_2&:=\{R\subseteq\s_1\t\s_2\,;\,R_1[p]\in\l_1,\,R_2[p]\in\l_2,
\,\forall
p\in\s_1\t\s_2\}\, ,
\end{split}
\]
ordered by set-inclusion, where
$R_1[(p_1,p_2)]:=\{s\in\s_1\,;\,(s,p_2)\in R\}$ and similarly,
$R_2[(p_1,p_2)]:=\{t\in\s_2\,;\,(p_1,t)\in R\}$.
\end{definition}

\begin{figure}[t]
\setlength{\unitlength}{1cm} \frame{
  \begin{picture}(12,6)(0,0)
    \put(-0.75,0){\includegraphics{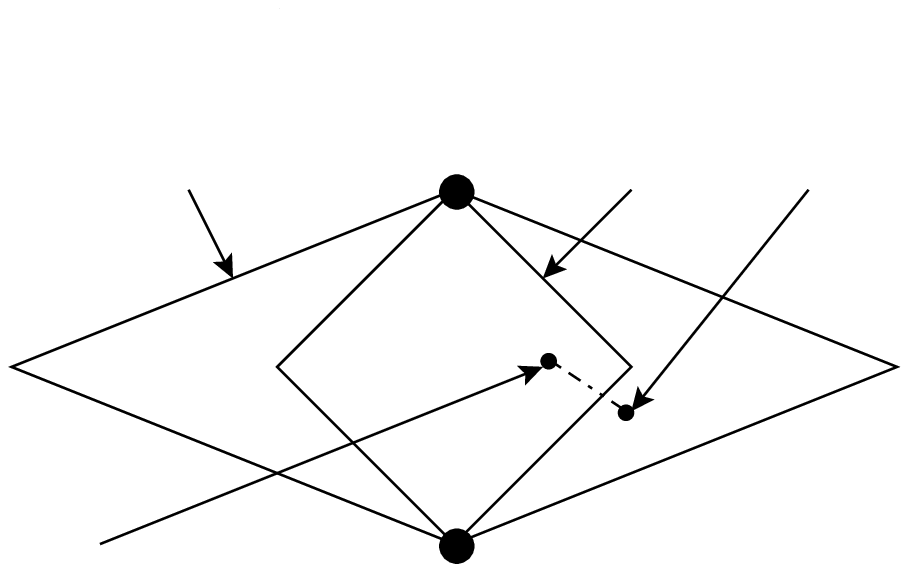}}
    \put(6,0.8){\makebox(0,0){$\l_1\aerts\l_2$}}
    \put(6,5.5){\makebox(0,0){$\l_1\otf\l_2$}}
    \put(3,5.3){\makebox(0,0){$\sep(\l_1,\l_2)$}}
    \put(8,5.4){\makebox(0,0){$\Sep_{_{T_1T_2}}(\l_1,\l_2)$}}
    \put(2.2,1){\makebox(0,0){$\l_1\circledast\l_2$}}
    \put(10,5.3){\makebox(0,0){$\l_1\ots\l_2$}}
  \end{picture}
}
\caption{$T_i=\aut(\l_i)$} \label{FigureS(L_1,L_2)}
\end{figure}

As a first result we find easily that for any
$T_i\subseteq\aut(\l_i)$, $\Sep_{_{T_1T_2}}(\l_1,\l_2)$, ordered
by set-inclusion, is a complete lattice, the bottom and top
elements of which are given by $\l_1\aerts\l_2$ and $\l_1\otf\l_2$
respectively (see \cite{Ischi:2004}, Theorem 2.13). The meet is
the set-intersection, hence $\Sep_{_{T_1T_2}}(\l_1,\l_2)$ is a
meet-sublattice of $\sep(\l_1,\l_2)$.

Moreover, suppose that if $\l_i\ne 2^{\s_i}$, then there are two
atoms, say $p$ and $q$, such that $p\vee q$ contains a third atom
(say $r$) and covers $p$, $q$ and $r$. Then,
$\l_1\aerts\l_2=\l_1\otf\l_2$ if and only if $\l_1=2^{\s_1}$ or
$\l_2=2^{\s_2}$ (see \cite{Ischi:2004}, Theorems 5.2 and 5.4).

The bottom element $\l_1\aerts\l_2$ is the {\it separated product}
of Aerts defined for ortholattices in \cite{Aerts:1982} (see
\cite{Ischi:2004}, Lemma 3.2). For atomistic lattices (not
complete) with $1$, $\l_1\aerts\l_2$ can be defined in a similar
way by taking only finite intersections. Then, it is isomorphic to
the {\it box product} $\l_1\square\l_2$ of Gr\"atzer and Wehrung
\cite{Graetzer/Wehrung:1999}, and if $\l_1$ and $\l_2$ are
moreover coatomistic, to the {\it lattice tensor product}
$\l_1\boxtimes\l_2$ (see \cite{Ischi:2004}, Theorem 3.8).

On the other hand, the top element $\l_1\otf\l_2$ is the
$\boxtimes-$tensor product of Golfin \cite{Golfin:handbook}, and
it is isomorphic to the tensor products of Chu
\cite{Barr:handbook} and Shmuely \cite{Shmuely:1974} (see
\cite{Ischi:2004}, Theorem 3.14). Let $\cC$ be the category of
complete join-semilattices with maps preserving arbitrary joins,
and $\cc$ the subcategory of $\cC$ defined by considering as
objects simple closure spaces. Let $\l_1$ and $\l_2$ be simple
closure spaces. Then there is a bimorphism
$f:\l_1\t\l_2\rightarrow\l_1\otf\l_2$, such that for any object
$\l$ of $\cC$ or $\cc$ and any bimorphism
$g:\l_1\t\l_2\rightarrow\l$, there is a unique arrow $h$ such that
the following diagram
$$\bfig
 \ptriangle|alm|/>`>`>/[\l_1\t\l_2`\l_1\otf\l_2`\l;f`g`!h]
 \efig$$
commutes (see \cite{Ischi:2004}, Theorem 3.20). For
join-semilattices and maps preserving finite joins, this is
exactly the definition of the {\it join-semilattice tensor
product} given by Fraser in \cite{Fraser:1976}. Hence, we can call
the top element the {\it complete join-semilattice tensor product}
or simply the tensor product in the category $\cc$.

Note that for any $T_i\subseteq\aut(\l_i)$,
$\Sep_{_{T_1T_2}}(\l_1,\l_2)$ can be defined as the set of all
simple closure spaces satisfying the above universal property with
respect not to all objects $\l$ and bimorphisms $g$, but with
respect to a given class of objects and bimorphisms (see
\cite{Ischi:2004}, Theorem 4.4). Therefore, it is natural to call
elements of $\sep(\l_1,\l_2)$  weak tensor products of $\l_1$ and
$\l_2$.
\subsection{Orthocomplemented weak tensor products}

If $\l_1$ and $\l_2$ are orthocomplemented simple closure spaces,
then the binary relation on $\s_1\t\s_2$, defined as
$(p_1,p_2)\#(q_1,q_2)\Leftrightarrow p_1\p q_1$ or $p_2\p q_2$,
induces an orthocomplementation of the separated product
$\l_1\aerts\l_2$. Coatoms have the form
$(p_1,p_2)^\#=p_1^\p\t\s_2\cup\s_1\t p_2^\p$. Our main result
states that the separated product is the only orthocomplemented
weak tensor product. More precisely, we have:

\begin{theorem}[\cite{Ischi:2004}, Theorem
8.6]\label{TheoremTheTheorem} Let $\l_1$, $\l_2$ be
orthocomplemented simple closure spaces with the covering
property, and let $\l\in\sep(\l_1,\l_2)$. Then $\l$ admits an
orthocomplementation if and only if $\l=\l_1\aerts\l_2$.
\end{theorem}

We outline the proof in case $\l_i$ are irreducible and $\l$ is
transitive, {\it i.e.} the action of $\aut(\l)$ on the set of
atoms of $\l$ is transitive (note that this is a consequence of
Axiom P4 if $T_1$ and $T_2$ act transitively on $\s_1$ and $\s_2$
respectively). First it follows easily from Axiom P3 that if $x_1$
is a coatom of $\l_1$ and $x_2$ is a coatom of $\l_2$, then
$X:=x_1\t\s_2\cup\s_1\t x_2$ is a coatom of $\l$. We prove that
all coatoms of $\l$ are of this form. Denote by
$':\l\rightarrow\l$ the orthocomplementation of $\l$. Then $X'$ is
an atom of $\l$, say $p$. Let $q$ be another atom. Since $\l$ is
transitive, there is an automorphism $u\in\aut(\l)$ such that
$u(p)=q$. Define $u':\l\rightarrow\l$ as $u'(a):=(u(a'))'$. Then
$u'$ is an automorphism of $\l$. Moreover,
$q'=u(p)'=u(p'')'=u'(X)$. Then the proof follows directly from

\begin{theorem}[\cite{Ischi:2004}, Theorem 7.5]
Let $\l_1$, $\l_2$ be simple closure spaces such that the join of
any two atoms contains a third atom. Let $\l\in\sep(\l_1,\l_2)$,
and let $u\in\aut(\l)$. Then there is a permutation $\sigma$ and
two isomorphisms $v_i:\l_i\rightarrow\l_{\sigma(i)}$ ($i=1,2$)
such that for all $p\in\s_1\t\s_2$, $u(p)_{\sigma(i)}=v_i(p_i)$.
\end{theorem}

The proof relies on the following remarks: Let $p$,
$q\in\s_1\t\s_2$. (i) By Axiom P2, if $p_1\ne q_1$ and $p_2\ne
q_2$, then $p\vee q$ does not contain a third atom. (ii) By Axiom
P3, if $p_1=q_1$, then $p\vee q=p_1\t (p_2\vee q_2)$, and the same
kind of equality holds for left lateral joins of atoms. As a
consequence, since $u$ preserves joins, $u(p_1\t\s_2)$ is either
of the form $q_1\t\s_2$ or of the form $\s_1\t q_2$, with $q_i$
atoms.

A similar result to Theorem \ref{TheoremTheTheorem} was obtained
in \cite{Ischi(rmp):2004} for $\l_i=\pro(\h_i)$ and with a set of
axioms weaker than those used here and in previous works
\cite{Aerts/Daubechies:1979,Pulmannova:1985,Watanabe:2003}.
\subsection{Weak tensor products with the covering property}

It was proved by Aerts in case $\l_1$ and $\l_2$ are
orthocomplemented simple closure spaces, that if $\l_1\aerts\l_2$
has the covering property or is orthomodular, then $\l_1=2^{\s_1}$
or $\l_2=2^{\s_2}$ (\cite{Aerts:1982}, or see \cite{Ischi:2004},
Theorem 9.1).

The same result holds for the top element $\l_1\otf\l_2$. More
precisely, assume that $\l_1$ and $\l_2$ have the covering
property and that if $\l_i\ne 2^{\s_i}$, then there are four atoms
$p$, $q$, $r$ and $s$ such that $p\vee q$ covers $p$, $q$, $r$ and
$s$. Then, $\l_1\otf\l_2$ has the covering property if and only if
$\l_1=2^{\s_1}$ or $\l_2=2^{\s_2}$ (see \cite{Ischi:2004}, Theorem
9.4).

We now give an example of a weak tensor product with the covering
property, which, as discussed in Section
\ref{SubSectionMissingProperties}, is a very natural model for
$\l_{sep}$. Let $\h_1$ and $\h_2$ be complex Hilbert spaces and
let $\l_1=\pro(\h_1)$ and $\l_2=\pro(\h_2)$ be the lattices of
closed subspaces. Let $V$ be a closed subspace of $\h_1\ot\h_2$.
Denote by $\prs[V]$ the set of atoms of $\pro(\h_1\ot\h_2)$
contained in $V$ and spanned by product vectors. Define
\[\l_1\ots\l_2:=\{\prs[V]\,\vert\,V\in\pro(\h_1\ot\h_2)\}\, . \]
Then $\l_1\ots\l_2\in\Sep_{_{T_1T_2}}(\l_1,\l_2)$ with
$(T_1,T_2)=(\unit(\h_1),\unit(\h_2))$, the group of automorphisms
induced by unitary maps (note that the same inclusion holds for
pairs of antiunitary maps, but not for
$(T_1,T_2)=(\aut(\l_1),\aut(\l_2))$). Moreover, $\l_1\ots\l_2$ is
different from the top and the bottom elements, $\l_1\ots\l_2$ is
coatomistic and has the covering property, but is not a
DAC-lattice (see \cite{Ischi:2004}, Theorem 10.4). As an example,
consider the case where $\h_1$ and $\h_2$ have finite dimensions.
Then, there is a bijection between anti-linear maps from $\h_1$ to
$\h_2$ and coatoms of $\l_1\ots\l_2$, namely
$A\mapsto\{p\t(A(p))^\p\,\vert\,p\in\s_1\}$.
\subsection{A second example}

Let $\l_1$ and $\l_2$ be coatomistic simple closure spaces. Define
$\l_1\circledast\l_2:=\{\cap\omega\,\vert\,\omega
\subseteq\s_\circledast'\}$,
with
\[\s_\circledast':=\{R\subsetneqq\s_1\t\s_2\,\vert\, R_1[p]\in
\s_1'\cup\{\s_1\}\ \mbox{and}\
R_2[p]\in\s_2'\cup\{\s_2\},\,\forall p\in\s_1\t\s_2\}\, ,\]
where $\s_i'$ denotes the set of coatoms of $\l_i$ (hence $R_i[p]$
is either a coatom or $\s_i$). Then
$\l_1\circledast\l_2\in\Sep_{_{T_1T_2}}(\l_1,\l_2)$ for
$T_i=\aut(\l_i)$ (see \cite{IschiSeal:2004}, Theorem 7.8).

Let $\cCas$ be the category of coatomistic simple closure spaces
such that for any two coatoms $x$ and $y$, and any two atoms $p$
and $q$, there is an atom $r$ and a coatom $z$ with $r\notin x\cup
y$ and $p,\,q\notin z$, with maps preserving arbitrary joins,
sending atoms to atoms or $0$, and with right adjoint sending
coatoms to coatoms or $1$. Then, $\cCas$ equipped with the
bifunctor $\circledast$ and the functor $\,^\op$ which sends a
lattice to its dual, is $\ast-$autonomous (see
\cite{IschiSeal:2004}, Theorem 5.5), hence a model for Girard's
linear logic \cite{Barr:1991}.

Note also that there is a bijection between $\cCas(\l_1,\l_2^\op)$
and $\s_\circledast'\cup \{1\}$, namely $f\mapsto\{p\t
f(p)\,\vert\,p\in\s_1\}$. Hence, for finite-dimensional Hilbert
spaces, we have
\[\pro(\h_1)\ots\pro(\h_2)\subseteq\pro(\h_1)\circledast\pro(\h_2)\,
.\]
Therefore, according to the discussion of Section
\ref{SubSectionMissingProperties},
$\pro(\h_1)\circledast\pro(\h_2)$ might, as well as
$\pro(\h_1)\ots\pro(\h_2)$, be a good candidate for $\l_{sep}$.
\section{Open questions}\label{SectionOpenQuestion}

Below $\h_1$ and $\h_2$ are complex Hilbert spaces.
\begin{enumerate}
\item[Q1] Let $\l_1=\pro(\h_1)$ and $\l_2=\pro(\h_2)$. Is the
statement of Theorem \ref{TheoremTheTheorem} true if we assume
Axioms P1, P2, P4 with $T_i=\unit(\h_i)$, and that the maps
$a_1\mapsto a_1\t \s_2$ and $a_2\mapsto \s_1\t a_2$ preserve
arbitrary joins?
\item[Q2] Is it always true that
$\pro(\h_1)\circledast\pro(\h_2)\ne\pro(\h_1)\otf\pro(\h_2)$?
\item[Q3] Is it possible to classify weak tensor products with the
covering property?
\item[Q4] Is there any theorem like:`` Let $\l$ be a coatomistic
simple closure space with the covering property. Suppose that $\l$
is not a DAC-lattice. If ..., then there are simple closure spaces
$\l_1\,\cdots\l_n$ which are DAC-lattices such that (up to an
isomorphism) $\l\in\sep(\l_1,\cdots,\l_n)$''?
\end{enumerate}
Note that a partial answer to Question Q3 can be found in
\cite{Ischi:2004} (see Theorem 9.6).
\bibliographystyle{abbrv}
\bibliography{DenverIschi2}
\end{document}